\documentclass[letterpaper]{article} 
\usepackage{aaai2026}  
\usepackage{times}  
\usepackage{helvet}  
\usepackage{courier}  
\usepackage[hyphens]{url}  
\usepackage{graphicx} 
\usepackage{natbib}  
\usepackage{caption} 
\usepackage{algorithm}
\usepackage{algorithmic}

\usepackage{newfloat}
\usepackage{listings}
\DeclareCaptionStyle{ruled}{labelfont=normalfont,labelsep=colon,strut=off} 
\floatstyle{ruled}
\newfloat{listing}{tb}{lst}{}
\floatname{listing}{Listing}
\usepackage{xcolor}
\newcommand{\answerYes}[1]{\textcolor{blue}{#1}} 
 
\newcommand{\answerNA}[1]{\textcolor{gray}{#1}}

\title{You See It, They Don't: An Exploratory Study of User-to-User Variation in Instagram Comments}
\author {
    Brahmani Nutakki\textsuperscript{\rm 1},
    Manon Lilott Kempermann\textsuperscript{\rm 1},
    Ingmar Weber\textsuperscript{\rm 1}
}
\affiliations {
    \textsuperscript{\rm 1}Saarland Informatics Campus, Saarland University, Germany\\
    bnutakki@cs.uni-saarland.de, make00009@stud.uni-saarland.de, iweber@cs.uni-saarland.de
}

\begin{document}

\maketitle

\noindent
\thanks{\textbf{This work has been accepted at the International AAAI Conference on Web and Social Media (ICWSM) 2026 as a short paper. Once available, please cite the peer-reviewed version.}}
\\

\begin{abstract}
In March 2025, Meta announced a new AI system to rank the order of the comments shown to Instagram users. With existing research showing how feed personalization systems can lead to increased polarization, the introduction of this new system raises similar questions. This paper presents a small-scale exploratory study examining whether the ranking system produces systematic differences in visible comments shown to different users, particularly for news-related content. Using four sock-puppet accounts varying in gender and political leaning, we collect visible comments on posts from ten news and ten non-news accounts. This collection is repeated twice from two VPN locations to assess location effects. We ask 1) how many visible comments vary across different users, 2) is this variation higher for news accounts than non-news accounts, and 3) can user-attributes like gender, political leaning, and location systematically explain the observed variation. Contrary to our expectations, we find that visible comments on news posts are less likely to vary across users than those on non-news posts. Variation is better explained by account metrics like comment and follower counts than by user attributes. These findings provide an initial glimpse into personalized comment ranking on Instagram and motivate larger, more systematic audits of how comment personalization may shape online discourse. To support further research, we provide the code to collect comments and the data upon request.
\end{abstract}


\section{Introduction}
\label{sec:introduction}

Over time, social media platforms have come to function as public squares, where users can engage in discussions on diverse topics. One such extremely popular platform is Instagram, with reports finding that the app is installed on almost 80\% of smartphones worldwide, amounting to 3 billion Monthly Active Users \cite{datareportal}. As early as 2016, Instagram revealed that an average of 95 million photos (and videos) were shared per day, a figure that has likely grown since \cite{instagram_daily_posts}. Yet, surveys indicate that only 20\% of users actively post on the platform \cite{insta_passive_users}, implying that most users interact primarily by liking, sharing, or commenting on existing posts.

Of the above metrics, comments emerge as the epicenter of dialogue and engagement on the platform: even posts from accounts with 1000-2000 followers receive at least 1-3 comments on average \cite{Instagram_benchmarks}. This ratio grows drastically, particularly with larger accounts, which can garner tens of thousands of comments. Previous studies have shown that attention-grabbing comments that are highly engaging are frequently negative and emotionally charged \cite{risch2020top-3a3, heraki2025analyzing-ac5}. These comments can reduce a post’s credibility \cite{naab2020comments-76c}, and a user's intent to share it \cite{boot2021processing-c6e}. Conversely, user-initiated corrections can also effectively counter misinformation if they include reliable sources \cite{seo2022if-c6e}.

In March 2025, Meta announced that an AI system was being used to personalize comments shown to Instagram users \cite{meta_announcement}. This system takes different input signals, such as the likelihood of a user to report, delete, reply, click on, or scroll past a comment, to decide how they should be ranked \cite{meta_announcement}. As of writing this paper, Instagram lets users choose among three settings for the comments: `For You', `Most recent', and `Meta verified', with `For You' being the default selection. Given existing literature on how social media's algorithmic personalization may contribute to increased polarization \cite{pournaki2025how-6a0, cinus2022effect-79f}, this announcement has raised similar questions. While there are works that question whether personalization is the primary driver for the observed increase in polarization \cite{garimella2017longterm-280}, and the long-standing debate on the impact of personalization \cite{conover2021political-e91, dahlgren2021critical-140}, recent intervention experiments identified causal links between them \cite{piccardi2025reranking-b0e}. Exposure to diverse political content can even trigger ``backfire effects" increasing polarization \cite{bail2018exposure-f33}. Prior research also indicates that comments appearing at the top can shape the narrative that follows \cite{naab2020comments-76c, zhangCharacterizingOnlinePublic2018}. Yet, compared with the extensive literature on feed ranking, there appears to be relatively limited research on comment ranking.

In this context, we wanted to understand how this new ranking system might influence narratives and potentially amplify polarizing perspectives. Given that this is a relatively new development and conducting large-scale sock-puppet audits is challenging, we begin with a small-scale exploratory study. For this study, our goal is to examine how visible comments vary across users viewing news and non-news content. Nearly 55.8\% of Instagram users use the app to stay up-to-date with news \cite{datareportal}, and news is often polarizing. Social media news comment sections have also been characterized as real-time barometers of public opinion \cite{hossain2024visual-8d9}, with research showing that users who are typically highly engaged with news are more likely to actively comment on the posts \cite{kalogeropoulos2017who-06a}. Given this, news acts as a natural starting point to assess whether user attributes, such as gender, political leaning, and location, affect the comments they see. To provide a baseline, we compare this against non-news content. Our research questions include \textbf{1)} To what extent do comments vary across users viewing the same post? \textbf{2)} Is user-user variation greater for news-related posts than non-news posts? and \textbf{3)} Are user attributes associated with systematic differences in the comments shown?

To do this, we analyze visible comments displayed to four sock-puppet users with different political leanings and gender profiles: Female Democrat, Female Republican, Male Democrat, and Male Republican. For each, we collect the visible comments (visible without scrolling, referred to as comments from now on) shown on the same posts from ten news accounts and ten non-news accounts. To test location effects, we repeat the collection from two VPN endpoints in New York State and Texas. We selected these two diverse US locations as most of the news accounts are US-based.

Contrary to our initial hypothesis, our results show that, on average, only 12\% of the comments differ across users. Comments on news posts are actually less likely to vary than those on non-news posts. Account metrics such as follower and comment counts are highly associated with variation but in opposite directions. By comparison, user attributes have only modest effects and do not appear to be the primary predictors.

To our knowledge, this is the first study to investigate personalized comment ranking on Instagram in this context. We hope this exploratory study highlights the need for more systematic research on this development. As platforms increasingly introduce personalization to boost engagement, understanding how these decisions shape discourse and their impact on an already fragmented society becomes vital. Both the code used to collect comments and the data itself are available upon request.

\section{Data Collection}
\label{sec:datacollection}

The data collection pipeline consists of three main steps: (1) creating sock-puppet user accounts, (2) collecting recent posts from selected Instagram accounts, and (3) collecting the top visible comments for each post as viewed by the sock-puppet accounts. The steps are described in detail in the subsections below.

\subsection{Sock-Puppet Accounts Creation}

To understand how comments might differ across different users, we created four new sock-puppet accounts using different email addresses. To maintain uniformity across accounts, we used the same date of birth and used gibberish usernames to avoid any username-specific inference by the platform. These accounts were set up with four personalities: Female Democrat, Female Republican, Male Democrat, and Male Republican. The bio was left empty, and the gender of the account was assigned in the settings, matching the gender assigned to the email address. For political-leaning preferences, we manually followed politicians from the Democratic and Republican parties for the algorithm to infer the account's preference. No other actions such as liking, sharing or commenting were performed to avoid unforeseen effects. Despite this limited interaction, the user and suggested feeds of the sock-puppet accounts showed personalization, with Republican accounts having predominantly Republican posts and the Democratic accounts following a similar pattern. The list of politicians followed can be seen in the appendix.

\subsection{Posts Collection}

We selected ten Instagram accounts belonging to news outlets across the political spectrum, using bias ratings from Media Bias/Fact Check \cite{MediaBiasFactCheck}. To compare these with non-news content, ten non-news accounts from different niches, such as sports, entertainment, food, and pets, were also selected (for both see Table \ref{tab:accountsdetails}). We ensured that the follower counts of news and non-news accounts collected roughly match to reduce unintended variability.

After identifying the accounts, we collected their ten most recent posts that were posted at least 24 hours before the point of data collection to ensure that their comment counts are saturated. While collecting posts that were not yet saturated could have provided more insights, we refrained from doing so to minimize uncertainty. More details about saturation can be found in Section \ref{sec:discussion}. In total, this resulted in 200 posts from twenty different news and non-news accounts.

\begin{table*}[!htb]
    \centering
    \begin{tabular}{c|c|c||c|c|c}
         News Account&  MBFC Bias&  \#followers&  Non-News Account&  Account Type& \#followers\\\hline
         MSNBC&  Left&  2.4M&  Peacock&  Entertainment& 2.4M\\
         Huffington Post&  Left&  3.3M&  Nytcooking&  Food& 4.6M\\
         CNN&  Center-Left&  21.7M&  Espn&  Sports& 28.4M\\
         Washington Post&  Center-Left&  7.3M&  Catloversclub&  Pets& 8M\\
         Forbes&  Center&  7.3M&  Thedogist&  Pets& 7.6M\\
         The Hill&  Center&  304K&  Thegradecricketer&  Sports& 323K\\
         Washington Times&  Center-Right&  120K&  Pbsfood&  Food& 156K\\
         New York Post&  Center-Right&  2M&  Hulu&  Entertainment& 2.6M\\
 Fox News& Right& 10.8M& Ladbible& Entertainment&15.2M\\
         Breitbart&  Right&  1.8M&  Accesshollywood&  Entertainment& 1.8M\\
    \end{tabular}
    \caption{This table provides the accounts used in the data collection process, along with other details. Follower counts are reported as of January 4, 2025.}
    \label{tab:accountsdetails}
\end{table*}

\subsection{Comments Collection}

Once the sock-puppet accounts and posts were available, we built a Selenium crawler to open each post and collect the caption, timestamp, and the number of comments. After logging in with each sock-puppet account, the crawler also collected the comments that were visible on each post without scrolling, along with their timestamps and the username of the commenter. We ran this collection process for every post under each sock-puppet account using two proxy locations: one in New York and one in Texas. The comments were collected at least 24 hours post the creation of the accounts to allow for personalization. This produced eight separate crawls in total—four accounts multiplied by two locations—each collecting top visible comments (usually 10-15) from the same set of 200 posts.

For the analysis, we limited attention to the top ten visible comments, since most posts contained at least ten. We collected only text-based comments, as the crawler could not reliably capture GIFs or hidden comments. Manual checks indicated that only about 30\% of posts contained GIFs in the top comments; among those, most had only one or two. We also removed generic automated comments such as “You can review or change your choices at any time in your cookie settings.” After filtering, we obtained around 930 comments from news accounts and 970 comments from non-news accounts for each crawl. After combining comments across all crawls, we obtained 980 and 1037 unique comments from news and non-news accounts, respectively.

\section{Empirical Analysis and Results}
\label{sec:methodology}

This section describes the approaches used and the insights they provide on the visibility of comments across different posts and different users.

\subsection{Descriptive Analysis}

As mentioned in Section \ref{sec:datacollection}, we collect comments from eight user crawls. To answer RQ1, we start by calculating the proportion of variation between each pair of crawls. For two crawls $A$ and $B$, and for each post $p$, let $C_A(p)$ be the set of top comments on $p$ collected in crawl $A$ and $C_B(p)$ for those collected in crawl $B$. We define the variation between the two crawls on post $p$, denoted by $V(p)$, as the symmetric difference between the two sets (comments that appear in exactly one crawl):

\begin{center}
    $V(p) = (C_A(p) \cup C_B(p)) - (C_A(p) \cap C_B(p))$
\end{center}

We then normalize by the total number of collected comments across the two crawls for the post $T(p) = |C_A(p)| + |C_B(p)|$ to obtain the per-post variation proportion as $V(p)/T(p)$. The overall average variation proportion between crawls $A$ and $B$ is then computed by averaging across all posts:

\begin{center}
    $\frac{1}{N}\sum_{p=1}^{N} V(p)/T(p)$
\end{center}
, where $N$ is the total number of posts across all accounts.

\begin{figure}[!htb]
    \centering
    \includegraphics[width=1\linewidth]{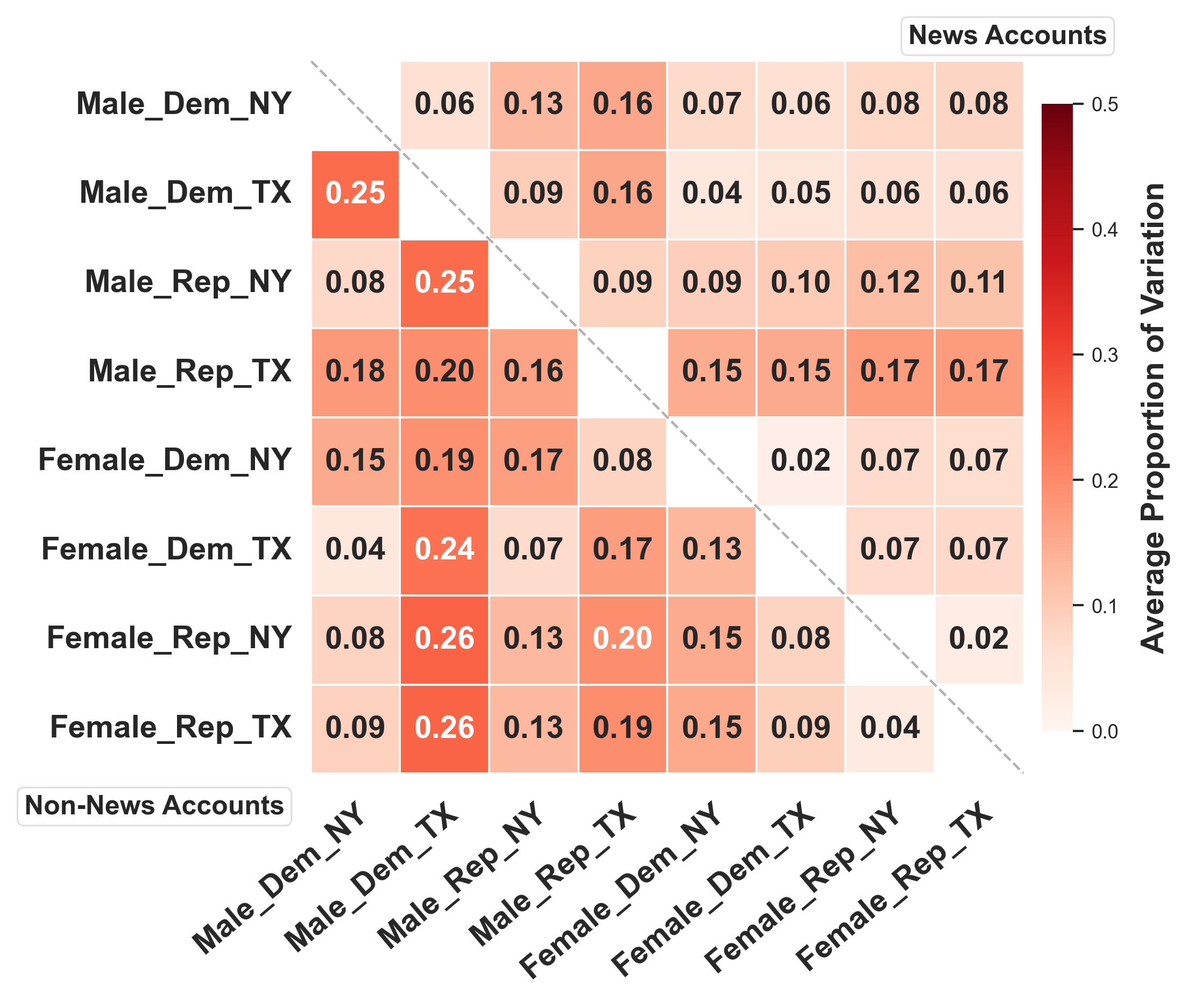}
    \caption{This heatmap shows the average variation proportion of comments across all posts between each pair of crawls. The upper triangle contains values for posts from News Accounts, and the lower triangle for those from Non-News Accounts.}
    \label{fig:heatmap_no_rank}
\end{figure}

We repeat this for each pair of crawls to characterize how variation proportions differ across each pair. The resulting heatmaps for news and non-news accounts are shown in Figure \ref{fig:heatmap_no_rank}. Overall, on average, only 12\% of comments vary between crawls. Comments on non-news accounts exhibit higher average variation than news accounts across crawls, suggesting that post type may be an important factor. While the heatmaps reveal some variation between crawls, the patterns are subtle and do not show strong structural variation.

\subsection{Post-level Regression Analysis}

To understand variations at a post level and answer RQ2 and RQ3, we fit a Bayesian generalized linear mixed model with a beta-binomial likelihood to model the variation observed for each post across a given crawl-pair comparison. Beta-binomial is chosen rather than binomial to account for over-dispersion. Each observation corresponds to one post and one crawl pair comparison; the outcome is variation $V(p)$ out of total trials $T(p)$ for post $p$, for all 28 crawl pairs ($C^{8}_{2}$). Fixed effects include non-directional labels for Location, Gender, and Leaning, derived from the two crawls being compared (e.g., comparing `Female\_Dem\_NY' with `Male\_Rep\_TX' produces the labels `Female\_Male', `Dem\_Rep,' and `NY\_TX'), along with comment and follower counts (log-transformed and standardized), and whether the post originated from a news or non-news account. To account for repeated observations from the same post, we added a random intercept for post ID. Model parameters and convergence statistics are included in the appendix.

\begin{figure}[!htb]
    \centering
    \includegraphics[width=1\linewidth]{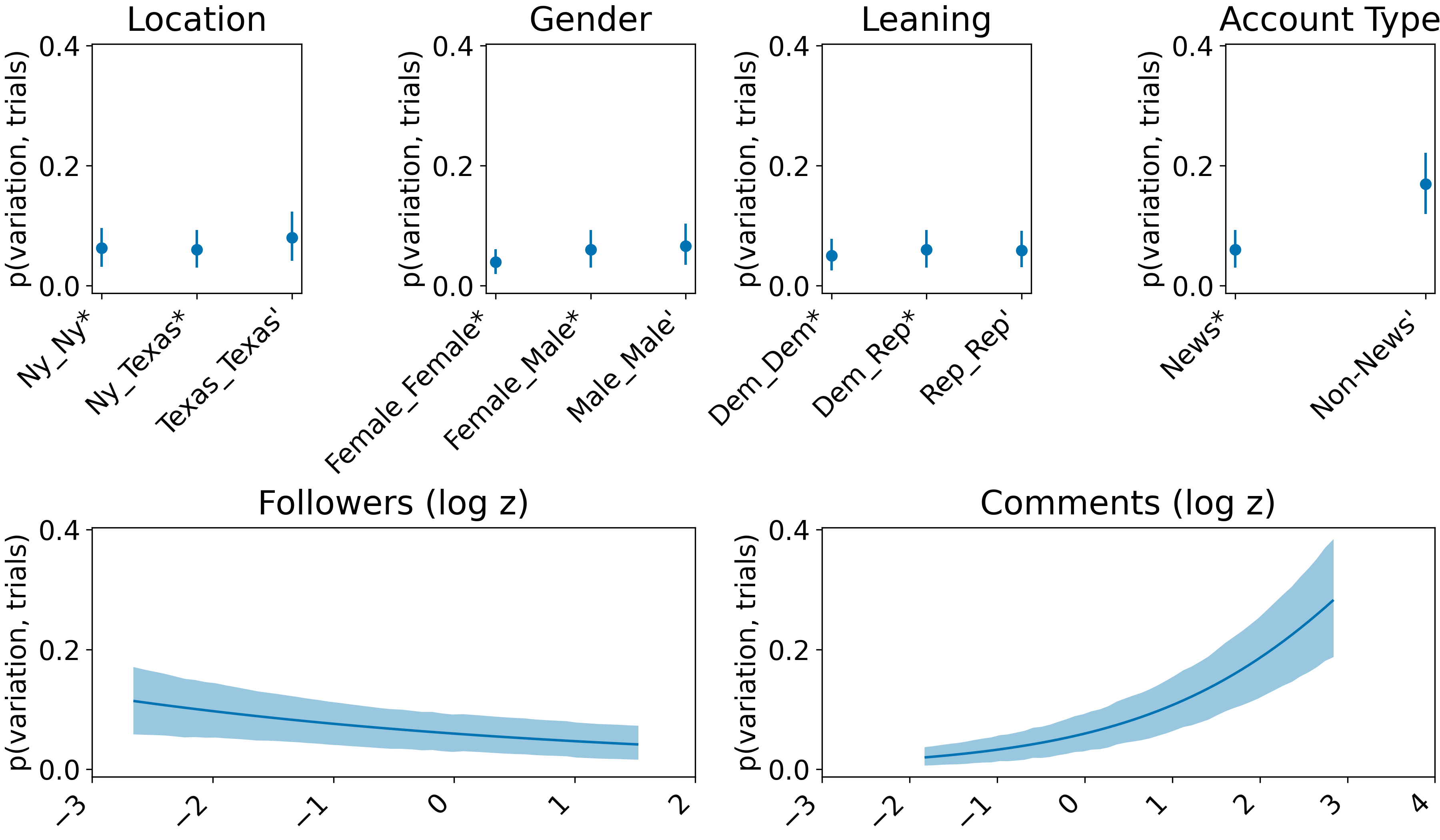}
    \caption{Posterior mean predictions (95\% HDI) from the beta-binomial regression across categorical and continuous covariates, holding other predictors constant. The categories with ' are the reference categories, and * indicates categories whose credible interval for the odds ratio excludes 1.}
    \label{fig:ppp_combined}
\end{figure}

The posterior mean prediction plot (see Figure \ref{fig:ppp_combined}) shows that account type and engagement metrics are more strongly associated with the predicted variation proportion. Holding other covariates constant, the model’s posterior mean variation proportion is higher for non-news posts (0.17) than for news posts (0.06). It also increases steadily as the comment count increases. In contrast, follower counts show a negative association. Location, Gender, and Political Leaning effects are comparatively small in absolute terms (on the order of only a few percentage points on the probability scale). For several categories of Location, Gender, and Political Leaning, the 95\% highest density intervals (HDI) for the odds ratio exclude 1, indicating evidence of association. However, the estimated magnitudes are small, suggesting limited practical significance, similar to the insights from the heatmaps. A forest plot of posterior odds ratios is provided in the appendix.

\subsection{Comment-level Regression Analysis}

To understand if the nature of comments is associated with variation, we ran additional analysis using posts from both news and non-news accounts. Each comment is labeled as `Supportive', `Against', or `Neutral' relative to the post caption using OpenAI API (Exact prompt can be found in the appendix). These annotations were validated by comparing them against 30 manually annotated comments, and no discrepancies were found. For each crawl pair, we tracked each unique comment and coded whether it appeared in both crawls. The outcome was a binary indicator for non-overlap, where 0 meant present in both crawls and 1 meant present in only one crawl, so the model estimates the probability that a comment is missing from one of the two crawls. We fit a Bayesian Bernoulli (logistic) model with the same sampling configuration as before. Fixed effects are the same as the above, along with the comment label.

The results show that supportive comments have a lower estimated probability of non-overlap than against (and neutral) comments, indicating they are more consistently present across crawls. The effects of the remaining variables mirror the previous findings. The corresponding probability plot is provided in the appendix, due to space constraints.

\section{Discussion and Conclusion}
\label{sec:discussion}

To the best of our knowledge, this is the first systematic study to investigate whether different users see different comments on Instagram posts following the introduction of the AI-driven comment ranking system.

Looking at the research questions outlined above, we find that overall variation is low: on average, 12\% of the comments differ between crawls (RQ1). Comments on news accounts exhibit less variation than those on non-news accounts (RQ2). Account metrics such as follower and comment counts are also associated with variation, but in opposite directions: higher comment counts indicate higher variation, while lower follower counts indicate higher variation. One possible interpretation is an exploit-explore trade-off \cite{Gisselbrecht_Denoyer_Gallinari_Lamprier_2015}: posts with many comments provide richer engagement signals, which may prompt the system to exploit these signals to personalize more. Whereas, smaller accounts might need more exploration and thus higher observed variation. While we do observe limited differences associated with location, gender, and political leaning (RQ3), given the current design, we can not establish their robustness.

To mitigate the effects of the rapidly evolving comments section, we only collect posts that are older than 24 hours. We chose this threshold by estimating the time needed to reach 95\% of comment count after 72 hours of posting (approximately 12–17 hours on average), by collecting comments from newly published posts and tracking their growth in five-hour intervals. Similarly, the data collection for the 200 posts in the dataset for all eight crawls was done with a short time window of 4-5 hours to avoid the addition of new comments or other unknown factors. A brief check showed that the number of comments that were added during these 4-5 hours was extremely low (less than 10 across all posts).

Various factors may be associated with the observed variations. Since our sock-puppet accounts were newly created, the observed limited variance could also be attributed to the algorithm's limited knowledge about the users. We only use four sock-puppet accounts, which also limits the robustness of our results. We also observed that the commenters mostly aligned with the political bias of the news account (e.g, predominantly right-leaning comments on posts from Breitbart), which also explains supportive comments having a low probability of being present in only one crawl. Choosing a more contentious domain where both pro- and anti- comments are prevalent would have provided more interesting results. Lastly, in this study, we only look at the presence of a comment, i.e, whether the comment appears across crawls, not ranking order. A preliminary analysis of common comments across crawls showed that ranking variation is substantial. The average variance proportion (considering difference in rank as variation) was 0.5 for news accounts and 0.7 for non-news accounts, indicating that the order of visibility differs largely across users.

Despite comments playing an important role in driving engagement and shaping narratives, there is surprisingly little work examining how comment-ranking systems influence what users see. Our work tries to explore the impact of the new comment-ranking system of Instagram by setting up a small-scale sock-puppet study. While the findings are modest, this study provides a starting framework for assessing how comment ranking could reinforce or amplify existing biases, and we hope that this study underscores the need for more such systematic research.

\section{Acknowledgments}

Ingmar Weber, Brahmani Nutakki, and Manon Kempermann are supported by funding from the Alexander von Humboldt Foundation and its founder, the Federal Ministry of Education and Research (Bundesministerium für Bildung und Forschung).

\bibliography{bibliography}

\section{Ethics Checklist}

\begin{enumerate}

\item For most authors...
\begin{enumerate}
    \item  Would answering this research question advance science without violating social contracts, such as violating privacy norms, perpetuating unfair profiling, exacerbating the socio-economic divide, or implying disrespect to societies or cultures?
    \answerYes{Yes}
  \item Do your main claims in the abstract and introduction accurately reflect the paper's contributions and scope?
    \answerYes{Yes}
   \item Do you clarify how the proposed methodological approach is appropriate for the claims made? 
    \answerYes{Yes}
   \item Do you clarify what are possible artifacts in the data used, given population-specific distributions?
    \answerYes{Yes, in the Discussion Section.}
  \item Did you describe the limitations of your work?
    \answerYes{Yes, in the Discussion Section.}
  \item Did you discuss any potential negative societal impacts of your work?
    \answerNA{NA}
      \item Did you discuss any potential misuse of your work?
    \answerNA{NA}
    \item Did you describe steps taken to prevent or mitigate potential negative outcomes of the research, such as data and model documentation, data anonymization, responsible release, access control, and the reproducibility of findings?
    \answerNA{NA}
  \item Have you read the ethics review guidelines and ensured that your paper conforms to them?
    \answerYes{Yes}
\end{enumerate}

\item Additionally, if your study involves hypotheses testing...
\begin{enumerate}
  \item Did you clearly state the assumptions underlying all theoretical results?
    \answerYes{Yes}
  \item Have you provided justifications for all theoretical results?
    \answerYes{Yes}
  \item Did you discuss competing hypotheses or theories that might challenge or complement your theoretical results?
    \answerYes{Yes}
  \item Have you considered alternative mechanisms or explanations that might account for the same outcomes observed in your study?
    \answerYes{Yes}
  \item Did you address potential biases or limitations in your theoretical framework?
    \answerYes{Yes, in the Discussion Section.}
  \item Have you related your theoretical results to the existing literature in social science?
    \answerYes{Yes}
  \item Did you discuss the implications of your theoretical results for policy, practice, or further research in the social science domain?
    \answerYes{Yes}
\end{enumerate}

\item Additionally, if you are including theoretical proofs...
\begin{enumerate}
  \item Did you state the full set of assumptions of all theoretical results?
    \answerNA{NA}
	\item Did you include complete proofs of all theoretical results?
    \answerNA{NA}
\end{enumerate}

\item Additionally, if you ran machine learning experiments...
\begin{enumerate}
  \item Did you include the code, data, and instructions needed to reproduce the main experimental results (either in the supplemental material or as a URL)?
    \answerYes{Yes, code and data will be provided upon request.}
  \item Did you specify all the training details (e.g., data splits, hyperparameters, how they were chosen)?
    \answerYes{Yes, in the appendix.}
     \item Did you report error bars (e.g., with respect to the random seed after running experiments multiple times)?
    \answerYes{Yes}
	\item Did you include the total amount of compute and the type of resources used (e.g., type of GPUs, internal cluster, or cloud provider)?
    \answerYes{The model does not require large computation.}
     \item Do you justify how the proposed evaluation is sufficient and appropriate to the claims made? 
    \answerYes{Yes}
     \item Do you discuss what is ``the cost`` of misclassification and fault (in)tolerance?
    \answerNA{NA}
  
\end{enumerate}

\item Additionally, if you are using existing assets (e.g., code, data, models) or curating/releasing new assets, \textbf{without compromising anonymity}...
\begin{enumerate}
  \item If your work uses existing assets, did you cite the creators?
    \answerNA{NA}
  \item Did you mention the license of the assets?
    \answerNA{NA}
  \item Did you include any new assets in the supplemental material or as a URL?
    \answerNA{NA}
  \item Did you discuss whether and how consent was obtained from people whose data you're using/curating?
    \answerNA{NA}
  \item Did you discuss whether the data you are using/curating contains personally identifiable information or offensive content?
    \answerNA{NA}
\item If you are curating or releasing new datasets, did you discuss how you intend to make your datasets FAIR?
\answerNA{NA}
\item If you are curating or releasing new datasets, did you create a Datasheet for the Dataset? 
\answerNA{NA}
\end{enumerate}

\item Additionally, if you used crowdsourcing or conducted research with human subjects, \textbf{without compromising anonymity}...
\begin{enumerate}
  \item Did you include the full text of instructions given to participants and screenshots?
    \answerNA{NA}
  \item Did you describe any potential participant risks, with mentions of Institutional Review Board (IRB) approvals?
    \answerNA{NA}
  \item Did you include the estimated hourly wage paid to participants and the total amount spent on participant compensation?
    \answerNA{NA}
   \item Did you discuss how data is stored, shared, and deidentified?
   \answerNA{NA}
\end{enumerate}

\end{enumerate}

\begin{appendix}

\section{Appendix}

\subsection{List of Politicians Followed by Sock-Puppet Accounts}

For the sock-puppet account to have a Democratic-leaning, we followed politicians including Former President Joe Biden (@joebiden), Former Vice President Kamala Harris (@kamalaharris), Former President Barack Obama (@barackobama), Representative Alexandria Ocasio-Cortez (@aoc), Senator Elizabeth Warren (@senwarren), Senator Bernie Sanders (@berniesanders) and Mayor Zohran Mamdani (@zohrankmamdani). For the Republican-leaning accounts, we followed President Donald Trump (@realdonaldtrump), Vice President J.D. Vance (@jdvance), Secretary of State Marco Rubio (@marcorubio), Secretary of Defense Pete Hegseth (@petehegseth), Governor Ron DeSantis (@rondesantis), Secretary of Health and Human Services Robert F. Kennedy Jr. (@seckennedy) and Director of National Intelligence Tulsi Gabbard (@tulsigabbard), among others (titles as of January 2025).

\subsection{Regression Models}

This subsection presents the model parameters, convergence diagnostics, and forest plots displaying the odds ratios (with 95\% HDIs) for the models described in Section~\ref{sec:methodology}.

\subsubsection{Model Parameters}

Unless otherwise specified, all models were fit using four Markov chains, each with 2000 posterior draws after tuning for 2000 steps at a target acceptance of 0.95. Results are reported at 95\% highest-density intervals (HDIs). A fixed random seed of 42 was used to ensure reproducibility.

\begin{figure}[!htb]
    \centering
    \includegraphics[scale=0.28]{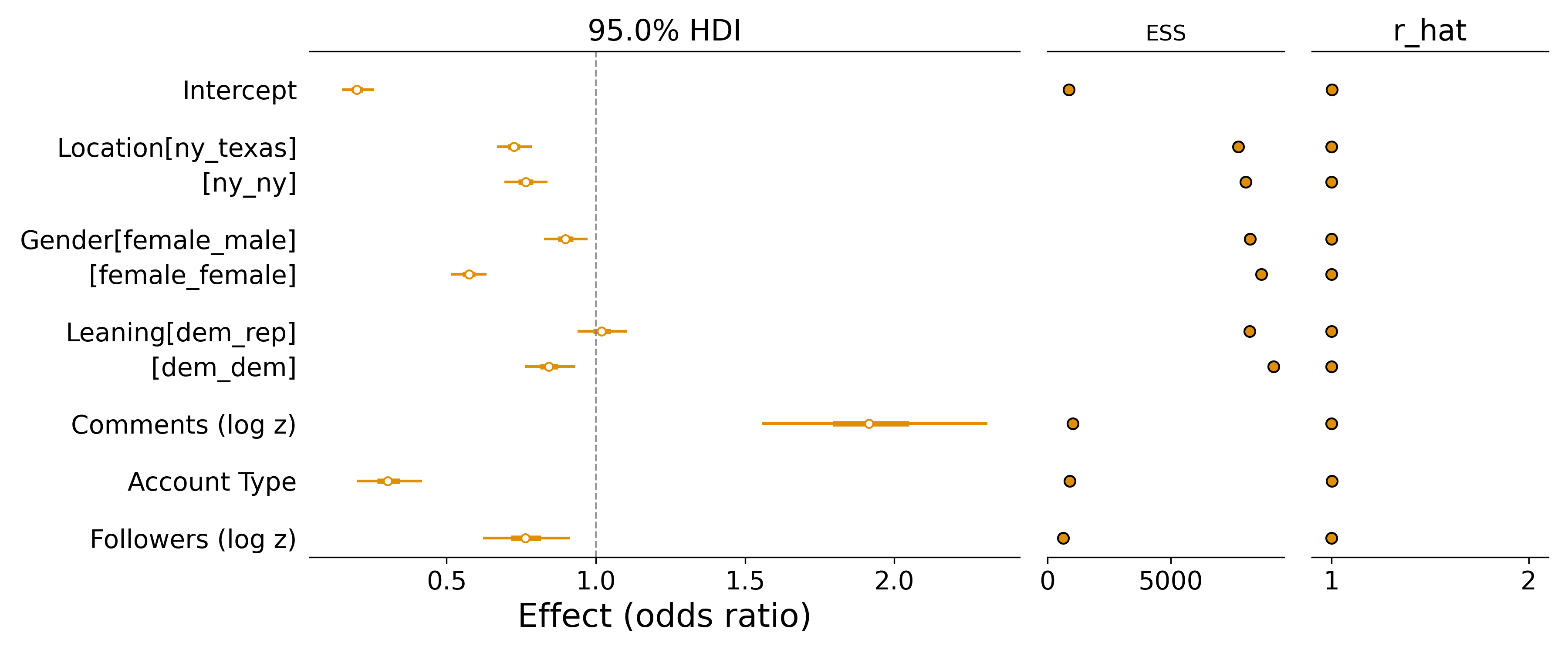}
    \caption{Forest plot of Bayesian logistic regression coefficients shown as odds ratios for Post-Level Regression. The dashed vertical line at 1 marks “no effect,” and the right panels report diagnostics for each parameter.}
\end{figure}

\begin{figure}[!htb]
    \centering
    \includegraphics[width=1\linewidth]{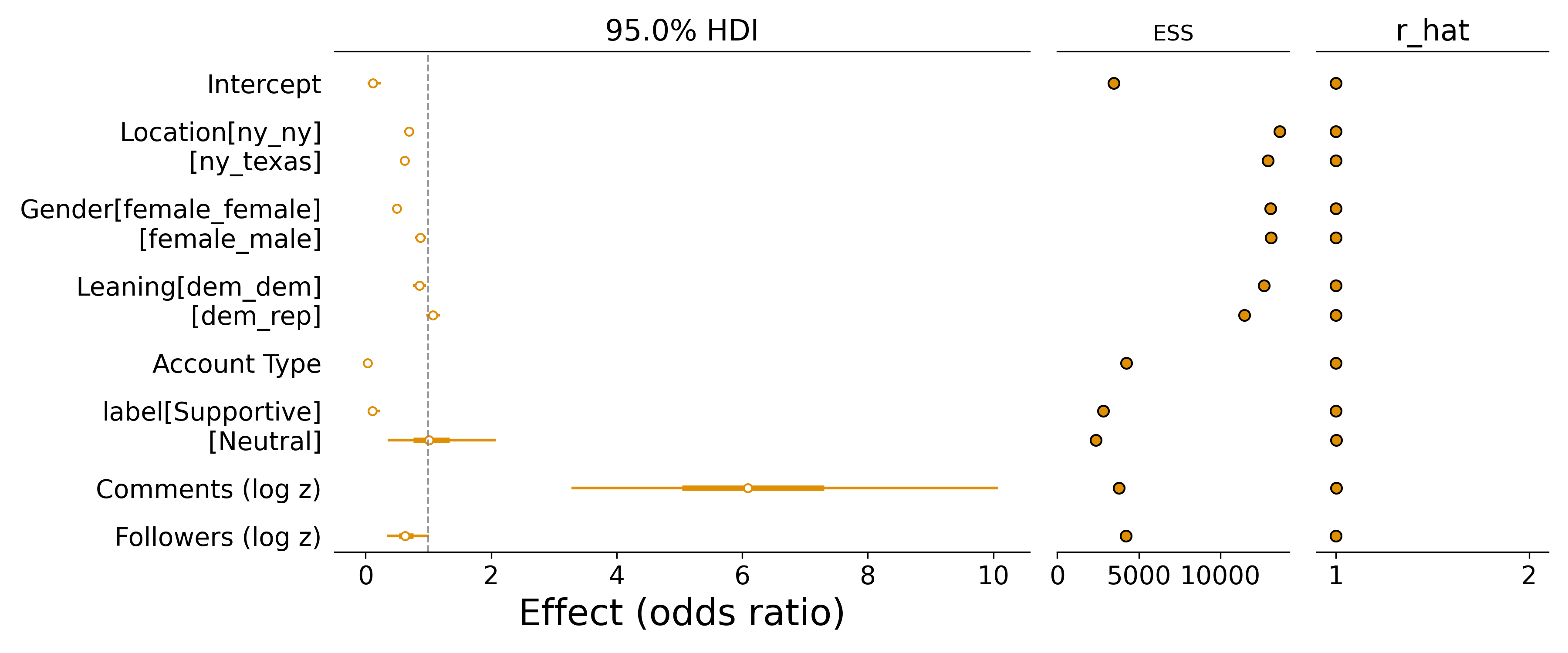}
    \caption{Forest plot of Bayesian logistic regression coefficients shown as odds ratios for Comment-Level Regression. The dashed vertical line at 1 marks “no effect,” and the right panels report diagnostics for each parameter.}
\end{figure}

\begin{figure}[!htb]
    \centering
    \includegraphics[width=1\linewidth]{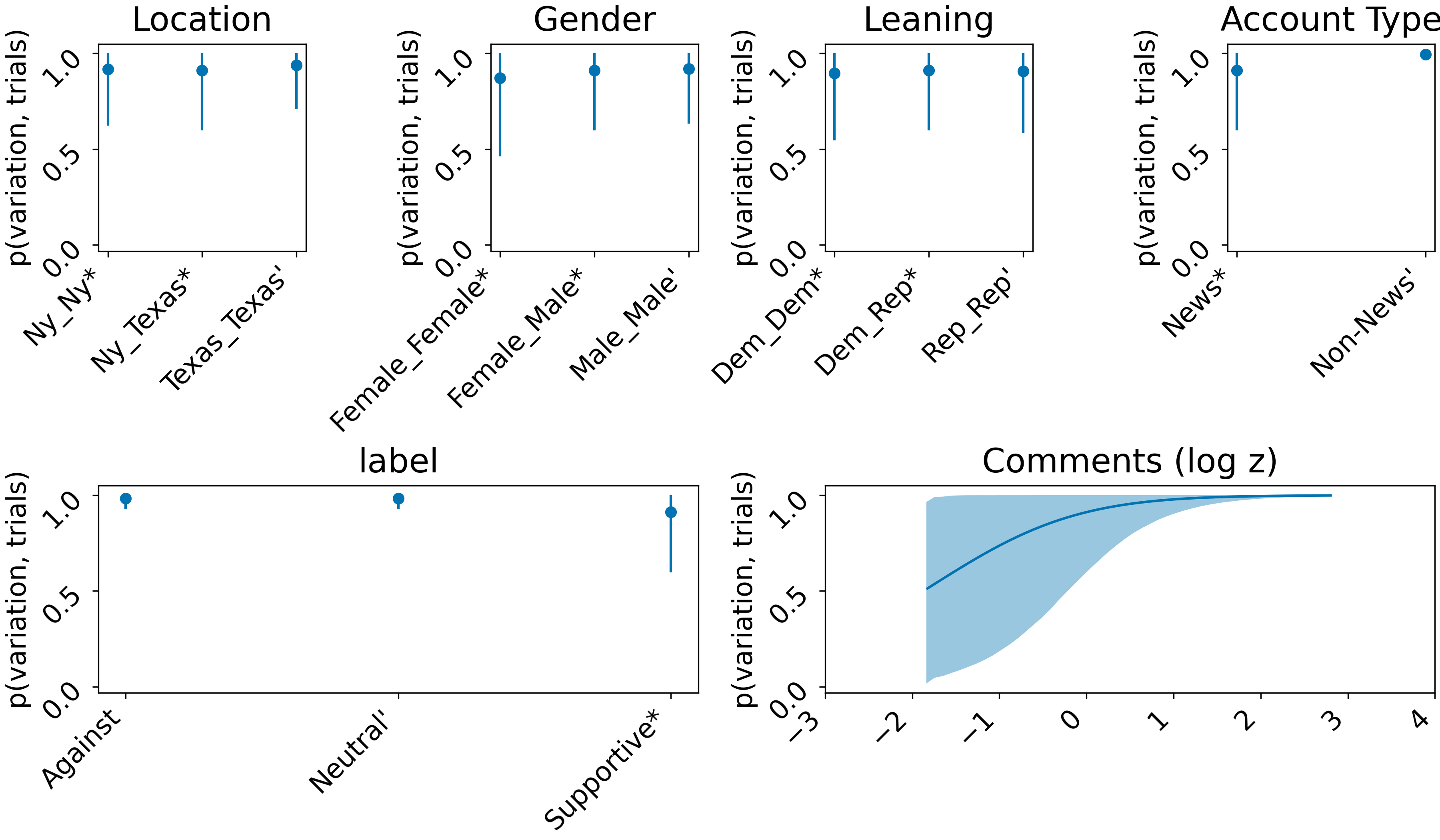}
    \caption{Posterior mean predictions for Comment-Level Regression across categorical and continuous covariates, holding other predictors constant. The categories with ' are the reference categories, and * indicates categories whose credible interval for the odds ratio excludes 1.}
\end{figure}

\subsection{Annotation of comments using OpenAI API}

The comments on each post from both news and non-news accounts were labeled `Supportive/Against/Neutral' relative to the post. To do this, we first manually annotated 30 randomly selected comments. We then used the following prompt to obtain labels for all the comments. No modifications were made to the default API settings. Finally, validating LLM annotations against the manually annotated comments showed 0 mismatches.

\subsubsection{Prompt}

You are a researcher annotating Instagram posts. Given the caption of an Instagram post and a comment on that post, you are trying to determine the stance of the commenter with respect to the caption.

Label the comment as one of the following categories:
\begin{itemize}
    \item Supportive: The comment expresses agreement, approval, or positive sentiment towards the content of the caption.
    \item Against: The comment expresses disagreement, disapproval, or negative sentiment towards the content of the caption.
    \item Neutral: The comment neither supports nor opposes the content of the caption; it may be factual or unrelated in tone.
\end{itemize}

\end{appendix}

\end{document}